\documentclass[epj]{svjour}
\usepackage{epsfig}
\usepackage[intlimits]{amsmath}
\usepackage{amssymb} \usepackage{amscd}
\renewcommand{\makeheadbox}{}
\newcommand{\Q}{\vec{Q}}
\newcommand{\kp}{k_{\|}}
\newcommand{\ks}{\vec{k}_{\perp}}
\newcommand{\et}[1]{E^{#1}_{\vec{k}}}
\newcommand{\kf}{k_F}
\newcommand{\en}{\Delta \epsilon_0}
\newcommand{\ef}{\epsilon_F}

\begin{document}
\title{Magnetic order and  transport in the heavy-fermion system
CeCu$_{6-x}$Au$_x$}
\author{H. v. L\"ohneysen\inst{1} \and A. Neubert\inst{1} \and  
A. Schr\"oder\inst{1} \and
O. Stockert\inst{1} \and U. Tutsch\inst{1} \and M. Loewenhaupt
\inst{2} \and
A. Rosch \inst{3}
\and P. W\"olfle \inst{3} 
}                     
%
%
\institute{Physikalisches Institut, Universit\"at Karlsruhe, D-76128 Karlsruhe,
Germany \and Institut f\"ur Angewandte Physik, 
Technische Universit\"at Dresden, 
D-01062 Dresden, Germany  \and 
Institut f\"ur Theorie der Kondensierten Materie, 
Universit\"at Karlsruhe, 
D-76128 Karlsruhe, Germany }
%
\date{}
%
   \abstract{ We report on extensive elastic neutron scattering to
     determine the wave vector of the magnetic order in
     CeCu$_{6-x}$Au$_x$ single crystals for $x > 0.1$.  For all values
     of $x$ investigated ($0.2$, $0.3$, $0.5$, $1.0$) we find
     long-range incommensurate antiferromagnetic order with an
     ordering vector $\Q \approx$ ($0.625$ $0$ $0.275$) for $x=0.2$,
     nearly unchanged for $x=0.3$, and $\Q \approx$ ($0.59$ $0$ $0$)
     for $x=0.5$, staying roughly the same for $x=1.0$.  In addition,
     short-range correlations are observed at $x=0.2$, reminiscent of
     those found previously for $x=0.1$.  The ordered magnetic moment
     is found to increase rapidly for small $x$, and more slowly for
     the larger $x$ values.  The increase of the specific-heat anomaly
     at the ordering temperature with $x$ is in qualitative accord
     with this behavior.  Finally, data of the electrical resistivity
     for current flow along the three crystallographic directions are
     presented, showing a clear signature of the magnetic order. A
     theoretical interpretation of the interplay of magnetic order and
     transport in terms of (i) the partial suppression of the Kondo
     effect by the staggered magnetization and (ii) the anisotropic
     band structure induced by the staggered field is shown to account
     well for the data, provided the ordering vector $\Q$ is close to
     $2 \kf$, where $\kf$ is a typical Fermi momentum.
} 
\maketitle

\section{Introduction}\label{section1}
Heavy-fermion systems exhibit a fascinating interplay of magnetically
ordered and non-magnetic groundstates and - in some systems -
superconductivity \cite{fulde88}.  It has been known for a decade that
the heavy-fermion compound CeCu$_6$ which does not show magnetic order
down to temperatures $T$ of at least 5\,mK
\cite{schuberth95,pollack95} exhibits long-range antiferromagnetic
order when alloyed with Au or Ag \cite{german88,gangopadhyay88}. From
the beginning, this has been explained \cite{german} as arising from
the interplay between onsite Kondo screening of the Ce magnetic
moments in a crystal-field split $^2$F$_{5/2}$ doublet groundstate
\cite{stroka} and the RKKY interaction between Ce moments. The latter
is favored by a weakening of the Kondo screening with increasing of
the interatomic spacing upon alloying with Au.  Indeed, the magnetic
order can be suppressed in CeCu$_{6-x}$Au$_x$ and a nonmagnetic
groundstate is recovered upon application of a sufficiently high
hydrostatic pressure \cite{german,bogenberger95}. The Kondo
temperature $T_K$ as estimated from the specific heat in large
magnetic fields $B = 6$\,T applied along the easy direction, decreases
monotonically from $T_K$ = $6.2$\,K ($x = 0$) to $4.6$\,K ($x = 0.5$)
\cite{schlager93,loehneysen96a}. In line with this $T_K$ decrease, the
onsite fluctuation rate as measured with inelastic neutron scattering
$\Gamma (T \rightarrow 0)$ which is attributed to the Kondo effect, is
smaller by a factor of $\sim 2$ in CeCu$_{5.5}$Au$_{0.5}$ than in pure
CeCu$_6$ \cite{stroka}.  Between $x = 0.1$ and $1$, the N\'eel
temperature $T_{\text{N}}$ rises from 0 to 2.3\,K and decreases
sharply beyond $x = 1$. For $x \leq 1$ Au occupies exclusively the
Cu(2) position in the orthorhombic CeCu$_6$ structure (Pnma)
\cite{loehneysen96}. The change of $d T_{\text{N}}/d x$ at $x = 1$
coincides with a subtle change within the orthorhombic structure: For
$x < 1$ the lattice parameters $a$ and $c$ increase while $b$
decreases with growing Au content, whereas for $x > 1$ all three
lattice parameters $a, b$ and $c$ increase. (We neglect the small
monoclinic distortion $(\approx 1.5^{\circ})$ of CeCu$_6$ occurring
below $T_S \approx$ 200\,K \cite{noda87}.  This structural transition
vanishes quickly with $x$, e.g. $T_S \approx$ 70\,K for $x = 0.1$
\cite{finsterbusch96}.)  Long-range antiferromagnetic order was
previously directly observed for $x = 0.5$, with incommensurate
reflections along the $a^*$ axis (we use the orthorhombic notation
throughout) indicating a magnetic ordering wave vector $\Q$ = (0.59 0
0) \cite{schroeder94}. At the critical concentration $x_c \approx 0.1$
for the appearance of long-range antiferromagnetic order, pronounced
deviations from Fermi-liquid behavior are observed in the
thermodynamic properties, i.e. specific heat and magnetization, and in
the electrical resistivity \cite{loehneysen94,loehneysen96}. The
critical fluctuations associated with this quantum critical point have
recently been identified \cite{stockert,schroeder}.

In this paper, we present a comprehensive study of the long-range
antiferromagnetism in CeCu$_{6-x}$Au$_x$.  We will use elastic neutron
scattering to characterize the magnetic ordering wave vector for
several concentrations $x=0.2, 0.3, 0.5$ and $1.0$. The rough estimate
of the ordered magnetic moment will be compared to that inferred from
the specific-heat anomaly.  Electrical resistivity measurements along
different directions reveal clear features attributed to the magnetic
order and represent an independent source of information on the
magnetic structure below $T_{\text{N}}$. We consider a
phenomenological theory of the transport of heavy quasiparticles in a
disordered lattice, which accounts well for the observed behavior. In
fact, the results allow to confirm the direction of the ordering $\Q$
vector. In addition they appear to indicate that the $\Q$ vector is
not too far from $2 \kf$, where $\kf$ is a typical Fermi momentum.

\begin{figure}[t]
  \begin{center}
   {\epsfclipon
   \epsfig{width=0.6 \linewidth,file=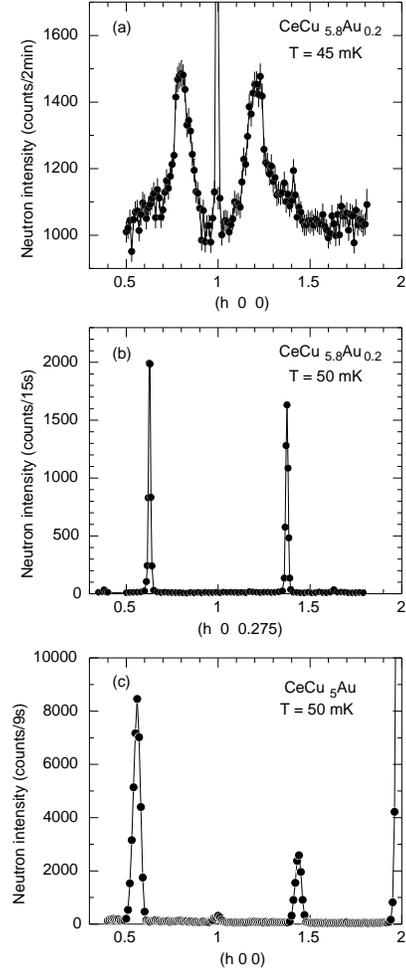}}
      \caption[]{a) and b): 
        Elastic scans in CeCu$_{5.8}$Au$_{0.2}$ along ($h$ 0 0) and ($h$ 0
        0.275) at temperatures $T \le 50$~mK well below $T_{\rm
          N}$~=~2.5\,K on IN 14 with a neutron energy $E$~=~2.7meV.
        The scans along the $a^*$ axis  reveal broad
        quasi-elastic structures indicating short-range correlations,
        while resolution-limited peaks are found at (0.625 0 0.275)
        and equivalent positions. \\
        c):
       Elastic scan along ($h$~0~0) in CeCu$_5$Au at $T$~=~50~mK
        ($T_{\rm N}$~=~2.3~K) on E4 ($E$~=~14~meV) showing
resolution-limited peaks with $\Q=$(0.56 0 0).}
    \label{fig1}
  \end{center}
\end{figure}

\section{Experimental}

All samples of this study were single crystals grown in a W crucible
with the Czochralski technique. The neutron scattering experiments
were performed at the Institut Laue-Langevin Grenoble, instrument
IN\,14 (for $x = 0.2)$ and the Hahn-Meitner-Institut Berlin,
instruments E\,4 and V\,2 ($x = 0.3, x = 1)$. The previous results for
$x = 0.5$ (as well as preliminary results for $x = 1$) were obtained
at NIST, instrument B\,9 \cite{schroeder94}. The quasi-adiabatic
heat-pulse technique was used for the measurement of the specific
heat.  The electrical resistivity $\rho(T)$ was measured on small
rectangularly shaped bars cut from the same crystals as used for the
neutron scattering. The standard four-probe technique was applied.
Because of the small sample size, the absolute $\rho$ values are
accurate only within 20\%.

\section{Results}
\subsection{Neutron scattering}

\begin{figure}[t]
  \begin{center}
   \epsfig{width=0.6 \linewidth,file=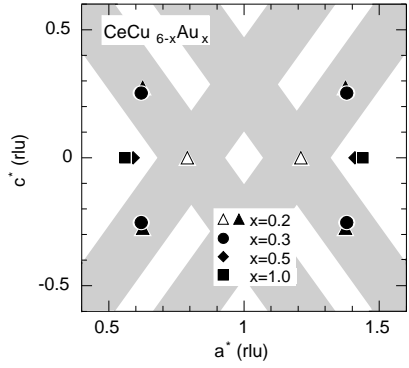}
      \caption{
        Position of the magnetic Bragg peaks ($x = 0.2$ -- $1.0$) in the
        reciprocal $ac$ plane in CeCu$_{6-x}$Au$_x$ (data for $x =
        0.5$ taken from \cite{schroeder94}).  The open symbols for $x
        = 0.2$ represent the short-range ordering peaks and the
        stripes indicate the dynamic correlations found for $x = 0.1$
        \cite{stockert}.  }
    \label{figPeaks}
  \end{center}
\end{figure}

Although the present paper focuses on the magnetically ordered
CeCu$_{6-x}$Au$_x$ alloys, we should mention at the outset that an
important issue is to identify the nature of the fluctuations
responsible for the nonmagnetic to magnetic transition at
the critical concentration $x_c=0.1$. We recently
identified two-dimensional short-range fluctuations in the dynamical
magnetic susceptibility for $x = 0.1$ from the observation of rod-like
structures in the reciprocal $ac$ plane \cite{stockert} (cf.
Fig.~\ref{figPeaks}). These dynamic correlations evolve for $x = 0.2$ into
short-range and long-range ordering peaks, both are located along
these rods. Fig.~\ref{fig1} shows results of elastic scans across magnetic
Bragg reflections taken at temperatures well below the ordering
temperature. For $x = 0.2$ we find the above mentioned short-range
magnetic order along the $a^*$ axis with a wave vector $\Q$ = (0.79
0 0) (Fig.~\ref{fig1}a). From the linewidth of the peaks, $\Delta q$ =
0.06\,r.l.u. (HWHM) in $a^*$, we deduce a correlation length of about
2.7 unit cells in the $a$ direction which is somewhat smaller than the
results previously reported \cite{rosch97} (there a factor of $1/(2 \pi)$
was omitted). This short-range order feature along the $a^*$ axis was
not observed for the $x $ = 0.3 alloy probably due to large background
compared to the expected magnetic intensity.
\begin{figure}[t]
  \begin{center}
   \epsfig{width=.95 \linewidth,file=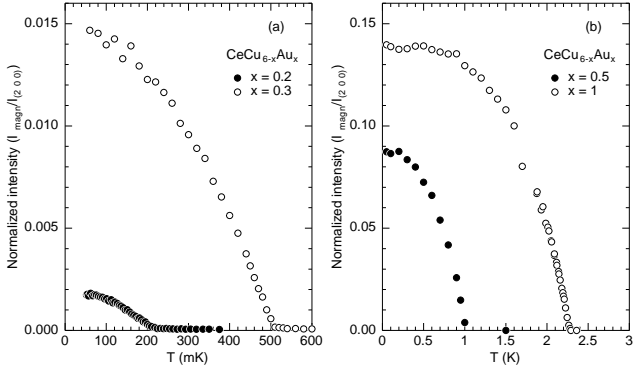}
      \caption{Temperature dependence of the intensity of corresponding
 Bragg peaks normalized to the (2 0 0) nuclear
reflection for $x=0.2$ ($\Q=$(0.625 0 0.275)) and
 $x=0.3$  ($\Q=$(0.62 0 0.253))in Fig. a and for $x=0.5$  ($\Q=$(0.59 0 0))
and $x=1.0$ ($\Q=$(0.56 0 0)) in Fig. b. 
 The intensity is proportional to the square of the staggered
magnetization $M_{\Q}(T)$. }
    \label{fig2}
  \end{center}
\end{figure}

In addition, we observe resolution-limited reflections for $x = 0.2$
in the $a^*c^*$ plane (Fig.~\ref{fig1}b), indicating long-range
magnetic order at $\Q$ = (0.625 0 0.275). Only minor changes in the
positions of the magnetic peaks are found for $x = 0.3$ with $\Q$ =
(0.62 0 0.253). In contrast, upon further Au doping for $x = 0.5$ the
magnetic order no longer appears off the $a^*$ axis, but
incommensurate order is observed along $a^*$ with $\Q$ = (0.59 0 0)
\cite{schroeder94} which is then roughly constant up to $x = 1$ ($\Q$
= (0.56 0 0 )).  Since these experiments on $x = 0.5$ and 1 were
performed long before the present ones on $x = 0.2$ and 0.3, we did
not take scans off the $a^*$ axis and therefore cannot exclude some
intensity at the positions out in the $a^*c^*$ plane. On the other
hand, powder measurements for $x = 0.5$ \cite{noda87} do suggest that
we did not miss an appreciable amount of magnetic intensity in our
investigations on the single crystals. Fig.~\ref{figPeaks} summarizes
the results obtained on the different CeCu$_{6-x}$Au$_x$ single
crystals. The rod-like feature in $S(q,\omega= 0.1$\,meV) of the alloy
at the critical concentration $x_c = 0.1$ and the positions of the
magnetic Bragg peaks are displayed in the reciprocal $ac$ plane.

Fig.~\ref{fig2} shows the intensity of selected magnetic Bragg peaks
as a function of temperature, one for each of the four investigated
concentrations $x = 0.2, 0.3, 0.5$ and $1.0$. The peak intensity as
normalized to the adjacent nuclear reflection, (2 0 0), is seen to
increase by a factor of 9 between $x = 0.2$ and 0.3, while it
increases more slowly between $x = 0.5$ and 1. It should be
mentioned that the intensity of the (2 0 0) nuclear peak depends
somewhat on the atomic positions inside the unit cell and is affected
by extinction since (2 0 0) is a strong peak. Therefore the normalized
intensity of the magnetic peaks is only a rough measure of the size of
the ordered magnetic moment.  Because of the change of the magnetic
structure and also of the scattering geometries between $x = 0.3$ and
0.5, the intensities of Figs.~\ref{fig2}a and \ref{fig2}b cannot be
compared directly. Assuming a sinusoidal modulation of the moments aligned
along $c$ we estimate an ordered magnetic moment $\mu$ of $0.1 \dots
0.15$\,$\mu_B$/Ce atom for $x = 0.2$. Under the same assumptions the
ordered moment for $x = 0.3$ is a factor of 3 larger. For $x = 0.5$
the estimate of $\mu \approx 1\,\mu_B$/Ce atom had been given previously
\cite{schroeder94}. Fig.~\ref{fig2}b shows that this value increases
only by small percentage for $x = 1$. These numbers should be compared
to 2.54\,$\mu_B$ for free Ce$^{3+}$ moments. 
For CeCu$_6$ the magnetization measured along the $c$ axis yields a
magnetic moment of 1.5\,$\mu_B$/Ce-atom in an applied field of 
40\,T \cite{sakihabara87}. Theoretically, the ordered magnetic
moment in a weakly interacting itinerant-electron model should depend
on the N\'eel temperatures as $\mu \propto T_{\text{N}}^{3/4}$
\cite{moriya} which gives a slower increase of the moment than
experimentally observed.

\begin{figure}[t]
  \begin{center}
   \epsfig{height=5cm,file=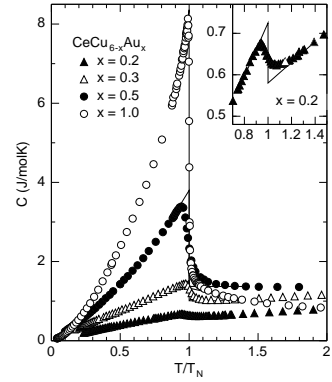}
      \caption{Specific heat for $x=0.2, 0.3, 0.5, 1.0$ as a function of
$T/T_{\text{N}}$. The lines correspond to a mean-field transition with the same
entropy contribution. The inset displays the specific heat for $x=0.2$.}
    \label{fig3}
  \end{center}
\end{figure}

 Although the ordering wave vector for $x=0.5$
and 1.0 as taken from ($h$ 0 0) scans is not changing very much,
the intensity of several reflections in the $a^*b^*$ plane off the
$a^*$ axis, not observed for $x=0.5$ is clearly not compatible with a
simple sine-modulated structure.
For $x = 1$ a complex magnetic ($B$,$T$) phase diagram has been observed, with
the magnetic structure found in zero applied magnetic field $B$ giving
way to a different structure in $B > 0$ via a first-order phase
transition \cite{paschke94}.  Neutron scattering studies in magnetic
fields are underway to examine the magnetic structure in these phases.

\subsection{Specific-heat anomaly at the ordering temperature}

The specific heat of the CeCu$_{6-x}$Au$_x$ single crystals has been
measured as reported in detail elsewhere
\cite{schlager93,loehneysen96a,paschke94}.  Here we focus on the
anomaly at $T_{\text{N}}$. Fig.~\ref{fig3} shows $C$ plotted vs.
reduced temperature $T/T_{\text{N}}$. As noted before
\cite{schlager93} the specific heat looks almost mean-field like for
$x =0.3$ and $ 0.5$. On the other hand it is more rounded for $x=0.2$
and more peaked for $x=1.0$.  There is a considerable contribution
arising from short-range ordering above $T_{\text{N}}$ as evidenced
from the tail of the anomaly.  The strong increase of the
specific-heat anomaly with $x$ qualitatively underscores the increase
of the ordered moment already inferred from the neutron scattering. A
more detailed analysis must await a microscopic model of the alloying
effect in CeCu$_{6-x}$Au$_x$ (see also section \ref{sectionTh}).  In a
crude analysis, we replace the observed anomaly at $T_{\text{N}}$ by a
mean-field discontinuity $\Delta C$ under the usual entropy -
conserving construction (cf. thin lines in Fig.~\ref{fig3}).  For an
ordered spin moment of $s = \frac{1}{2}$ one expects $\Delta C_{MF} =
1.5 R$ \cite{baker90}. We find the following values $r = \Delta
C/\Delta C_{MF}$ for $x = 0.2$, 0.3, 0.5 and 1: $r$ = 0.016, 0.04,
0.2, and 0.55, respectively.  One should keep in mind that even for $x
= 1$ a considerable Kondo effect is operative indicated by a sizable
$\gamma$ = 0.64\,J/molK$^2$ at 0.1\,K in zero magnetic field
\cite{paschke94}.

\begin{figure}[t]
  \begin{center}
   \epsfig{width=1. \linewidth,file=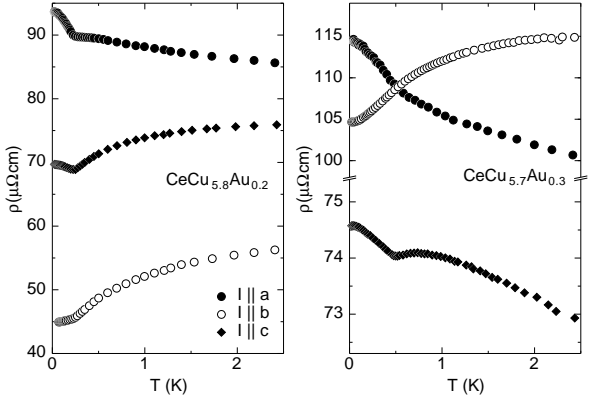}
    \caption{Resistivity of CeCu$_{5.8}$Au$_{0.2}$ and CeCu$_{5.7}$Au$_{0.3}$
      along the three crystallographic axes. Along the $a$ and $c$ direction a
sharp rise of the resistivity is observed at the N\'eel temperature
($T_{\text{N}} \approx 0.25$\,K for $x=0.2$ and $T_{\text{N}}\approx 0.51$\,K 
 for x=0.3).}
    \label{res1.fig}
  \end{center}
\end{figure}
\subsection{Resistivity}

Fig.~\ref{res1.fig} gives an overview over the $T$ dependence of the
electrical resistivity $\rho$ for $x = 0.2$ and 0.3. For both
concentrations, $\rho (T)$ shows a kink at the N\'eel temperature
$T_{\text{N}}$ and increases below $T_{\text{N}}$ for current direction $I$ parallel to
$a$ and $c$ while $\rho_b(T)$ for $I \parallel b$ continues to
decrease towards lower temperatures. For both concentrations
$\rho_a(T)$ has a negative temperature coefficient throughout the $T$
range investigated, and exhibits the largest magnitude compared to
$\rho_b$ and $\rho_c$ below $T_{\text{N}}$.

These findings suggest that below the N\'eel temperature
quasiparticle scattering properties are changed. This can happen in several
different ways as will be discussed in section~\ref{sectionTh}.
 We mention that resistivity data for $x = 0.15$ where
only $\rho_a$ and $\rho_b$ were measured \cite{tutsch} fit nicely into this 
picture, i.e. an increase of $\rho_a(T)$ below $T_{\text{N}}$ and a decrease of 
$\rho_b(T)$
is observed.

\begin{figure}[t]
  \begin{center}
   \epsfig{width=1. \linewidth,file=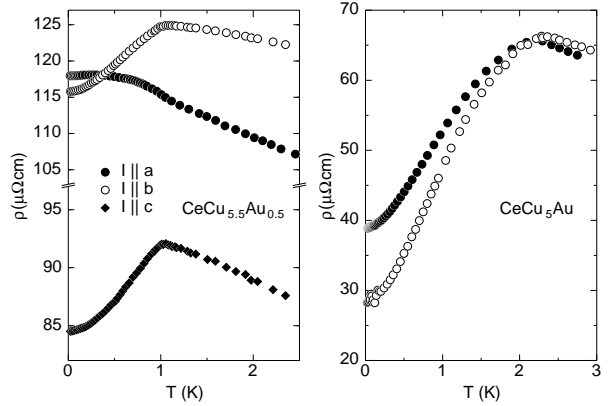}
    \caption{Resistivity of CeCu$_{5.5}$Au$_{0.5}$ and CeCu$_{5}$Au$_{1}$. 
      Along the $a$-axis a sharp rise of the resistivity is observed at
      the N\'eel temperature for $x=0.5$ ($T_{\text{N}} \approx 1.0$\,K). For
      $x=1$ ($T_{\text{N}} \approx 2.3$\,K) the resistivity decreases with
      temperature both in $a$ and $b$ direction, however the decrease
      along a seems to be smaller.}
    \label{res2.fig}
  \end{center}
\end{figure}

Fig.~\ref{res2.fig} shows $\rho(T)$ for $x = 0.5$ and 1. While for $x
= 0.5$ $\rho_a(T)$ again exhibits a (weak) kink at $T_{\text{N}}$ and increases
faster towards low $T$, $\rho_b$ and $\rho_c$ reveal a rather sharp
maximum at $T_{\text{N}}$. Again, these features can be explained by referring
to the magnetic structure. For
this concentration, the magnetic ordering vector lies on the $a^*$
axis hence only this direction shows an increase of $\rho(T)$ below
$T_{\text{N}}$.

For $x = 1$ (Fig.~\ref{res2.fig}b) we observe a maximum of $\rho(T)$
at $T_{\text{N}}$ both for $I \parallel a$ and $I \parallel b$. However, the
decrease of $\rho_a(T)$ below $T_{\text{N}}$ is slower than that of
$\rho_b(T)$.  It is important to note that even for $x = 1$, a very
large residual linear specific-heat coefficient $\gamma =
0.64$\,J/molK$^2$ is observed in the magnetically ordered state
\cite{paschke94}.

As a final result, we plot in Fig.~\ref{resist} the residual
resistivity $\rho_0$ for the three current directions as a function of
Au concentration $x$. As already apparent from the data of
Fig.~\ref{res1.fig} and \ref{res2.fig}, $\rho_{0,a}$ is largest throughout
the concentration range. The maximum of $\rho_{0,a}$ at $x \approx 0.5$
reflects the large structural disorder although the system is
magnetically homogeneous as evidenced from the resolution-limited
Bragg peaks.

\section{Interplay
  of magnetic order and transport} \label{sectionTh}
As a theoretical model
for CeCu$_{6-x}$Au$_x$ we envisage a conduction band of heavy
quasiparticles generated by the Kondo effect at the Ce ions.  Their
effective mass $m$ and Fermi energy $\ef$ is controlled by the Kondo
temperature $T_K$, i.e. $m \sim (T_F/T_K) m_{\text{band}}$, $\ef \sim
T_K$, there $T_F$ and $m_{\text{band}}$ are the Fermi temperature and
bare mass of the weakly interacting conduction electrons.  The
substitution of Au for Ce in CeCu$_6$ is known to take place at a
special Cu site next to each of the four Ce atoms in the unit cell (see
section \ref{section1}). The overall effect of alloying on the heavy
fermion liquid is to lower the Kondo temperature mildly (up to a
factor $\sim 2$ for $x=1$), as inferred from the $\gamma$ coefficient
of the specific heat and the quasi-elastic linewidth ($\sim
T_K$) in neutron scattering.  It is not known at present,
whether this must be interpreted as a
lowering of the characteristic energy of the coherent state of the
Kondo lattice forming at low temperatures, caused by the lattice
distortions and the associated (probable) lowering of the conduction
electron density of states at the Fermi energy and the weakening of
the f-d hybridization, or as a local lowering of the Kondo temperature
at those Ce ions next to a Au atom.  In any
case it will be useful to distinguish the single-ion regime above the
so-called coherence temperature $T_0$, defined by the maximum in the
resistivity as a function of temperature, from the lattice-coherent state
below $T_0$.  

The most prominent consequence of coherence is that for
a regular lattice of Kondo ions, i.e. in the stoichiometric compounds
CeCu$_6$ and CeCu$_5$Au, the resistivity should tend to zero in the
limit $T\to 0$.
\begin{figure}[t]
  \begin{center}
   \epsfig{width=0.6 \linewidth,file=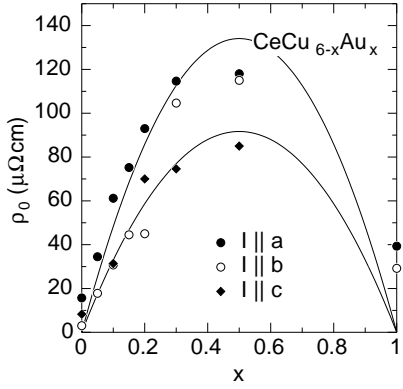}
      \caption{Residual resistivity along the $a$, $b$ and $c$ axis
      as a function of doping $x$. The solid lines denote the Nordheim
relation  $\rho \propto x(1-x)$.}
    \label{resist}
  \end{center}
\end{figure}
This is seen in Fig.~\ref{resist} where the residual resistivity is
plotted versus Au concentration $x$.  The fact that for $x=1$ the
residual resistivity $\rho_0$ is not quite zero, but assumes minimum
values of 30-40$\mu\Omega$cm, is likely due to remaining lattice
defects. The data points are roughly consistent with the law $\rho_0
\sim x (1-x)$. The maximal values for $\rho_0$ are of the order of the
unitarity limit, i.e.  correspond to a mean free path of the order of
the Fermi wavelength. This suggests that the Au impurities act as
strong scatterers, probably because they change the Kondo state of the
nearby Ce ion in a subtle way. We recall that the Kondo temperature in
the doped systems is only a factor two or so smaller than that of
CeCu$_6$, so we have no indication for a drastic suppression of the
Kondo effect by the Au impurities.  Nonetheless, the additional
potential scattering caused by the Au impurities may change the
f-level occupation on the Ce ions sufficiently to induce a relevant
deviation of the Kondo phase shift from the value $\pi/2$. In any
case, it is reasonable to assume that the observed strong increase of
$\rho_0$ with Au concentration is related to the Kondo effect in the
Ce ions.

This suggests the following scenario: at any finite temperature not
too far below $T_K$ the magnetic moments at the Ce ions are not
completely quenched by the Kondo effect. The residual moments interact
and may form an ordered magnetic state. Even in the fully developed
magnetically ordered state at $T=0$ the local magnetic field created
by the ordered spin configuration may be sufficiently small such that
a reduced Kondo effect remains. For a range of values of $T_K \gtrsim
T_{\text{N}}$ the Kondo effect and magnetic order may thus coexist. An
alternative scenario would be that of a spin density wave of the heavy
quasiparticles. While the latter model may have difficulties
to explain the rather large  magnetic moments experimentally observed at
$x=0.5$ and $x=1.0$, it can not be ruled out at present.

We consider now the effect of magnetic order on the resistivity.
Within the first scenario the onset of magnetic order would be
associated with a partial suppression of the Kondo effect by the local
magnetic field generated by the ordered moments.
The effect of a magnetic field $H$ on the resistivity $\rho_K$ of
a single $S=1/2$ Kondo ion at $T=0$ can be exactly expressed in terms
of the Kondo impurity magnetization $M_K$ as \cite{andrei}
\begin{eqnarray}
 \rho_K=\rho_K(0) \cos^2 \left(\frac{ \pi M_{K}}{g \mu_B}\right)
\end{eqnarray}
We expect this relation to hold approximately at finite
temperatures $T_0\lesssim T \lesssim T_K$ as well.
Taking the magnetization to be proportional to the ordered moment
$M_\Q(T)$, we obtain the following  estimate of a first effect
of the magnetic order on the resistivity of the Kondo alloys
in the single-ion regime
\begin{eqnarray}
  \label{rho1F}
  \rho(T,M_\Q(T))=\rho(T,M_\Q=0)  
\cos^2 \left(\frac{ \alpha M_{K}}{g \mu_B}\right)
\end{eqnarray}
where $\alpha$ is a coefficient of order unity. This will tend to
reduce all components of the resistivity tensor equally strongly. The
initial (isotropic) decrease below $T_{\text{N}}$ according to (\ref{rho1F}), 
\begin{eqnarray}
  \label{rho1}
   \frac{\delta \rho^{(1)}}{\rho_0} \approx - \frac{1}{2} \alpha^2
\left(\frac{M_{\Q}(T)}{g \mu_B}\right)^2,
\end{eqnarray}
is proportional to $M_\Q^2(T)$ and hence linear in $(T_{\text{N}}-T)$ near
$T_{\text{N}}$. For the experimentally determined temperature dependence of
$M_{\Q}(T)^2$ see Fig.~\ref{fig2}.

Secondly, the periodically modulated static spin structure in the
magnetically ordered state will give rise to a change in the
conduction-electron band structure. This is independent of whether the
magnetism is of an itinerant or a localized nature. Roughly speaking,
the additional periodic structure will give rise to enhanced
back-scattering for quasiparticles moving in the direction of the
magnetic wave vector $\Q$, hence increasing the resistivity components
along $\Q$. This is born out by the data of Fig.~\ref{res1.fig}, which
indeed show an increase of $\rho$ along the $a$ and $c$ directions, in
accordance with the direction of $\Q \approx (0.625~0 ~0.275)$ for
$x=0.2$  while $\rho$ increases only in the $a$ direction for $x=0.5$ with
$\Q\approx(0.59~ 0~ 0)$.

We will sketch a model calculation of this effect in the following.
The static magnetization associated with the magnetic order acts like
an additional (spin-dependent) periodic potential, thus changing the
band structure. As this change is small it can be calculated in
degenerate perturbation theory.  The band structure affects both the
quasiparticle energy $\epsilon_{\bf k}$ and the transport relaxation
time $\tau_{\bf k}$.  For a quantitative theory of this effect a
detailed knowledge of the band structure, the magnetic ordering, the
interactions of the staggered moment with the electrons and the
momentum dependence of $\tau_{\bf k}$ is necessary. The qualitative
effect, however, can easily be calculated, e.g. for a spherical Fermi
surface assuming isotropic scattering.  The staggered magnetization
$M_{\bf Q}(T)$ induces a spin-dependent periodic potential of wave
vector $\Q$.  It is now important to distinguish three different
scenarios, depending on whether the ordering wave vector ${\bf Q}$ is
larger, smaller or of the order of the size of the Fermi sphere with
radius $\kf$.  For $Q \gg 2 k_F$ the resistivity is practically not
affected by the short-range periodic potential. For $Q<2 k_F$ belts of
band-gaps are opened at the Fermi surface for $\epsilon_{\bf k}\approx
\epsilon_{{\bf k}\pm{\bf Q}}$.  The magnitude of those energy-gaps is
proportional to $M_{\bf Q}(T)$.


The model is described by the following Hamiltonian 
\begin{equation}
\label{hamilton}
H_0=\sum_{{\bf k},\sigma} \epsilon_{\bf k} c_{{\bf k}\sigma}^\dagger
c_{{\bf k}\sigma}+
\Delta \sum_{\bf k} \left( c_{{\bf Q}/2+{\bf k}\downarrow}^\dagger
c_{-{\bf Q}/2+{\bf k}\uparrow}+h.c.\right).
\end{equation}
${\bf Q}$ is the ordering wave vector, $\epsilon_{\bf k}$ is the
energy spectrum in the absence of the magnetic order.  We have assumed
a sinusoidal variation of the magnetization directed along the spin
quantization axis of the electrons. $\Delta$ measures the strength of
the effective potential and is proportional to the staggered
magnetization $M_{\bf Q}(T)$ measured in neutron scattering. The
energy eigenvalues of $H_0$ are given by ($\hbar=1$)
\begin{eqnarray}
\et{\pm}&=&
\frac{\epsilon_{\Q/2+\vec{k}}+\epsilon_{\Q/2-\vec{k}}}{2} \pm 
\sqrt{\left(
\frac{\epsilon_{\Q/2+\vec{k}}-\epsilon_{\Q/2-\vec{k}}}{2}\right)^2+\Delta^2}
\nonumber\\
&=&\en+\kp^2/2m +\ks^2/2m \pm \sqrt{ (v_0 \kp)^2+\Delta^2}  \label{band}
\end{eqnarray}
We will use a quadratic band structure $\epsilon_{\vec{k}}=k^2/2 m$ throughout
the paper, as no information is available on the complicated band structure
of  CeCu$_{6}$. The qualitative conclusions drawn from this
model will nonetheless be correct, although we expect that quantitative
changes will result from a realistic band structure.
As we are considering a heavy-fermion system we
expect $m$ to be very large, with the Fermi energy
$k_F^2/2 m \sim T_K$ ($\sim 6$\,K for CeCu$_6$)
 taking a rather low value. The variables
$\kp$ and $\ks$ are
the components of the momentum parallel and perpendicular to $\Q$ and $\kp$
is confined to the first (magnetic) Brillouin zone,
$-Q/2\le \kp \le Q/2$.
$v_0=Q/2m$ is the velocity at $\vec{Q}/2$ parallel to $\Q$ and $\en=
(\Q/2)^2/2m-\mu$ is negative (positive) for  $\Q< 2 \kf$ ($\Q> 2 \kf$),
where $\mu \equiv \kf^2/2m$ is the chemical potential.

In the following we will investigate within a simple Boltzmann transport
picture the
interplay of impurity scattering and the change of the band structure
due to the static magnetic order.
We will not consider inelastic processes like the scattering of electrons
from magnetic fluctuations, which are probably negligible at some distance
from the transition and at the low temperatures considered.
The change of the band structure affects
both the scattering rate of the electrons and the velocities of the fermions
at the Fermi surface. The change of the chemical potential due to $\Delta$
can be neglected as it is proportional to 
$ (\Delta/\ef)^2 \ln[ \ef/(\Delta+\en)]$ and therefore small 
compared to the other effects discussed below.

The transport scattering rate $1/\tau_{\vec{k}}$ near the Fermi surface
due to impurity scattering is given by 
\begin{eqnarray}
  \label{eq:1}
  \frac{1}{\tau_{\vec{k}'}} =\sum_{i=\pm} \int d^2 \ks d\kp 
\delta(\et{i}) (1-\cos \phi_{k k'}) W_{k k'}
\end{eqnarray}
where $\phi_{k k'}$ is the angle between $\vec{v}^{\pm}_{\vec{k}}=d
\et{\pm}/ d \vec{k}$ and $\vec{v}^{\pm}_{\vec{k}'}$.  For simplicity
we consider only s-wave scattering ($W_{k k'}=const.$), for which the
cosine drops out due to $\cos \phi+ \cos (\pi-\phi)=0$.  The
integration on $\kp$ yields a constant value for $0\le \kp \le
\kp^{\text{max}}$ where $\kp^{\text{max}}=Q/2$ for $i=-$ and
$\kp^{\text{max}}<Q/2$ for $i=+$, in the case that $Q<2 \kf$. (In the
opposite case $Q>2 \kf$ the contributions stem from
$\kp^{\text{max}}\le \kp \le Q/2$ and $i=-$ and $\et{+}$ is not
occupied.)  From the condition $\et{\pm}(\ks=0,\kp)=0$ we obtain
$\kp^{\text{max}}=\sqrt{\kf^2+(Q/2)^2-2 \sqrt{\kf^2 (Q/2)^2+m^2
    \Delta^2}} \approx \sqrt{(\kf-(Q/2))^2-m^2 \Delta^2/(\kf (Q/2))}$
with $\et{+}(0,\kp^{\text{max}})=0$ for $Q<2 \kf$ and
$\et{-}(0,\kp^{\text{max}})=0$ for $Q>2 \kf$. For $Q > 2 \kf$ only
$\et{-}$ gives a contribution to the scattering rate,
$1/\tau_{\vec{k}}\propto Q/2-\kp^{\text{max}}$. For $Q<2 \kf$ we
obtain from the lower band always the same contribution $\propto Q/2$
while from the upper band one has to add a contribution $\propto
\kp^{\text{max}}$. Combining this we find that the relative change of
the scattering rate due to the opening of a gap is given by
\begin{eqnarray}
  \label{scatter}
 \delta \left(\frac{1}{\tau_{\vec{k}}}\right) / \left(\frac{1}{\tau_{\vec{k}}}
\right)=& &  \nonumber\\
&&\makebox[-2cm][c]{} \frac{1}{\kf}
 \left( \frac{Q}{2}-\kf \mp \text{Re}
 \sqrt{\left( \frac{Q}{2}-\kf\right)^2-
\frac{m^2 \Delta^2}{\kf Q/2}}\right)
\end{eqnarray}
where the $+$ ($-$) is valid for $Q<2 \kf$ ($Q>2 \kf$). $\text{Re}$
denotes the real part, taking into account that the band gives no
contribution, if it is not occupied. For $Q<2 \kf$ the scattering rate
is {\em reduced} by $1-Q/(2 \kf)$ for $\Delta < |\en|$ and by
$\Delta^2/(8 \ef (Q/2)^2/(2m)) Q/(2 \kf -Q)$ for $\Delta > |\en|$.
For $Q>2 \kf$ the scattering rates are {\em increased} by the same
amount.  As the change of the scattering rate is approximately of the
order of $(\Delta/\ef)^2 (Q/(Q-2 \kf))$ large effects are expected
only for $Q \approx 2 \kf$.

The conductivity
due to impurity scattering is given by 
\begin{eqnarray}
\sigma_{\alpha \beta}&=&-e^2 \sum_{i=\pm}\int v^i_{\alpha} v^i_{\beta}
f'(\et{i}) \tau_{\bf k} d^3{\bf k}/(2 \pi)^3 \\
&\approx& e^2 \tau(\Delta)
  \sum_{i=\pm} \int 
\frac{\partial^2 \et{i}}{\partial k_{\alpha} \partial k_{\beta}} f(\et{i}) 
\label{mass}
\end{eqnarray}
where ${\bf v}^i=\partial \et{i}/\partial {\bf k}$ is the
velocity of the quasiparticles, $f(\epsilon_{\bf {k}})$ is the Fermi
function and $\tau_{\bf k}$ the (elastic) transport scattering time of the
quasiparticles.  

Let us first consider the components of $\sigma$ perpendicular to
$\Q$, $\sigma_{\perp}$. Within the model considered, the perpendicular
components of the inverse effective mass tensor $\partial^2 \et{i}/
\partial k_{\alpha} \partial k_{\beta}=1/m$ are unchanged by the
magnetic order. The conductivity is given by the Drude result
$\sigma_{\perp}=e^2 n \tau/m$ with, however, the proper relaxation
time $\tau$ changed by the gap according to Eq.~(\ref{scatter}). This yields
a
relative change of the resistivity $\rho_\perp$ induced by the
magnetic order, which is in leading order in $\Delta$ 
\begin{eqnarray}
  \label{rho2}
  \frac{\delta \rho^{(2)}_{\perp}}{\delta_0}=-\frac{1}{4} 
\frac{\kf^2}{Q(\kf-Q/2)} 
\left(\frac{\Delta}{\ef}\right)^2.
\end{eqnarray}
This estimate holds provided that $|\kf-Q/2|>m \Delta/\kf$. It is seen
that the conductivity is increased (decreased) for $Q<2 \kf$ ($Q>2
\kf$) by an amount proportional to the  squared amplitude of the magnetic order
$M_{\vec{Q}}(T)$.  Except for small $(\kf-Q/2)$ the effect is
of order $(T_{\text{N}}/T_K)^2$ where $T_{\text{N}}$ and $T_K$ are the
N\'eel and Kondo temperatures, respectively.

The conductivity component parallel to $\vec{Q}$ is mainly affected by
the change in the effective mass tensor
\begin{eqnarray}
  \label{massF}
  \frac{\partial \et{\pm}}{\partial \kp^2}=\frac{1}{m} \pm 
\frac{v_0^2 \Delta^2}{((v_0 \kp)^2+\Delta^2)^{3/2}}.
\end{eqnarray}
Performing the integration on $\ks$ in (\ref{mass}) one finds 
\begin{equation}
  \label{sigma2}
  \Delta \sigma^{(3)}_{\|}=-e^2 \tau m v_0^2 \Delta^2 \int 
\frac{d \kp}{(2 \pi)^2}
\frac{ T \ln [(1+e^{X_+})/(1+e^{X_-})]}{[(v_0 \kp)^2+\Delta^2]^{3/2}}
\end{equation}
where $X_{\pm}=-E^{\mp}(\kp,\ks=0)/T$. For $Q\lesssim 2 \kf$ 
($\en \lesssim 0$) and 
$\Delta \ll |\en|$ or else $\Delta \ll T$, one finds for the change in 
resistivity
\begin{equation} 
  \label{rho3}
  \frac{\delta \rho^{(3)}_{\|}}{\rho_0} =
\left(1- \frac{3 \pi}{2} \frac{v_0}{v_F} \frac{\Delta}{\ef} f(\en)
\right)^{-1} 
\approx   \frac{3 \pi}{2} \frac{v_0}{v_F} \frac{\Delta}{\ef} f(\en).
\end{equation}
This contribution has to be added to the one induced by the change in
the relaxation rate, $\delta \rho_{\|}^{(2)}/\rho_0=\delta
\rho_{\perp}^{(2)}/\rho_0$ which was found to be quadratic in $\Delta$,
and hence is smaller.

On top of the contributions (\ref{rho2}) and (\ref{rho3}) we have to add
the contribution found in (\ref{rho1}),
which is  in principle comparable in magnitude to (\ref{rho2}).

The theoretical estimates derived above allow to interpret the data
shown in Fig.~\ref{res1.fig} and \ref{res2.fig} in a satisfactory way.
For all Au concentrations shown, the resistivity increases below $T_{\text{N}}$
with decreasing temperature relative to the (extrapolated) background
only for those directions of current flow with a finite projection
onto the ordering vector $\Q$ (the effect being the larger, the larger
the projection). This is in accordance with our result that only for
those directions the dominant term linear in $\Delta$, given by
(\ref{rho3}), contributes. It remains to explain the temperature
dependence of this effect, which appears to be rather more linear than
square root.
\begin{figure}[t]
  \begin{center}
   \epsfig{width=0.8 \linewidth,file=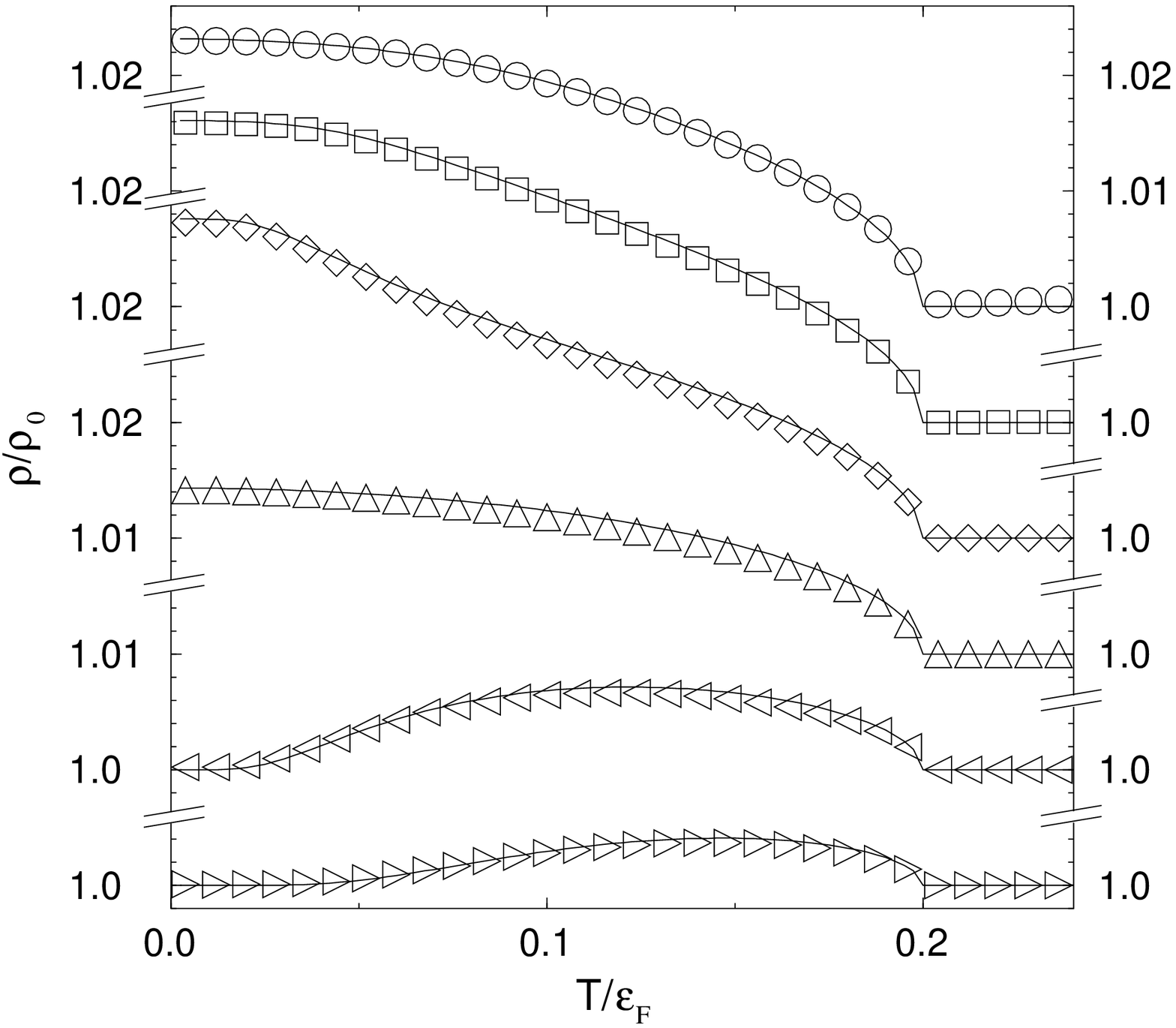}
      \caption{Increase of the re\-si\-sti\-vity $\rho^{(3)}_{\|}$ due
to band structure effects for the model described in Eqn.~(\ref{hamilton}) 
for $T_{\text{N}}=0.2 \ef$. The gap $\Delta(T)$ is assumed to 
be proportional to $M_{\Q}(T)$ derived from Fig.~\ref{fig2}: 
$\Delta(T)/\ef = 0.15 (T_{\text{N}}/\ef) \sqrt{1-(T/T_{\text{N}})^2}$. The result is given for
various ordering vectors from top to bottom: $Q/2=0.8 \kf, 0.9\kf, 0.95\kf, \kf, 1.05\kf, 1.1\kf$. For clarity we have plotted the curves with a constant
offset. The points are the result of a numerical evaluation of (\ref{mass}),
the solid lines show our approximating formula (\ref{rho3}).}
    \label{resNumerik}
  \end{center}
\end{figure}
However, although (\ref{rho3}) appears to give a square-root
dependence on ($T_{\text{N}}-T$) due to the linearity in $\Delta
\propto M_{\Q}(T)$, a numerical evaluation shows
(Fig.~\ref{resNumerik}) that for $\Q$ sufficiently close to $2 \kf$
and $T_{\text{N}}$ not too small compared to $\ef$ the temperature
dependence of the Fermi function $f(\en)$ in the relevant temperature
regime tends to straighten the square-root towards linear behavior
 over a wide range in
temperature.  This would indicate that for both $x=0.2, 0.3 $ and
$x=0.5, 1.0$, $\Q$ is close to spanning the distance between sections
of the Fermi surface with opposite Fermi velocities ($\Q \sim 1.8
\kf-2\kf$).  A fit of the results for $\delta \rho^{(3)}$ (\ref{rho3})
to the data is not easy due to the unknown background and the additional
isotropic contribution from $\rho^{(1)}$ and $\rho^{(2)}$. We find
roughly for the ratio $\Delta/\ef \approx 0.1 (T_{\text{N}}/T_K)
(M_{\Q}(T)/M_{\Q}(0))$.

Let us now turn to the components of $\rho$ perpendicular to $\Q$.  In
this case only the contributions $\delta \rho^{(1)}$ and $\delta
\rho^{(2)}$ survive, which are both quadratic in $M_{\Q}(T)$. Of
these, $\delta \rho^{(1)}$ is probably the dominant one, considering
the small value of $\Delta/\ef$. Using the values of the low
temperature staggered magnetization estimated from the neutron
scattering data, $M_{\Q}(0)/\mu_B\approx 0.1, 0.3, 1$ for $x=0.2, 0.3$ and
$1.0$ we find a drop in resistivity from $T_{\text{N}}$ to $T=0$ of relative
magnitude $\delta \rho^{(1)}(T=0)/\rho(T_{\text{N}})\sim 0.001, 0.01, 0.1$
taking $\alpha=1$ and $g=2 \mu_{\text{sat}}/\mu_B \approx 2.8$ in
(\ref{rho1}).  This compares well with the experimental data for which
an additional drop of $\rho_b$ towards lower temperatures, on top of
the decrease caused by the formation of coherence, is not noticeable
for $x=0.2$ and $0.3$, whereas for $x=0.5$ the relative change is
roughly $\delta \rho/\rho \sim 0.1$.  For the highest concentration
$x=1$, the predicted drop $\delta \rho^{(1)}/\rho\sim 0.15$ accounts
for part of the total drop, the main part being due to coherence
effects.  For $x=0.2$, the resistivity $\rho_b$ is seen to rise for
decreasing temperature with respect to an extrapolated background,
probably due to admixtures from components parallel to $\Q$. Another
possibility is that $\Q$ may be slightly larger than $2 \kf$ and
$\rho^{(2)}$ taken from (\ref{rho2}) is positive. However, this
is difficult to decide as long as the band structure is unknown and the
temperature dependence in the absence of magnetic order is not fully
understood.

In a more general view, the resistivity data show clear signs of the
existence of three characteristic temperatures, the Kondo temperature
$T_K$ ($\sim$ Fermi energy of the heavy quasiparticles), the
coherence temperature $T_0$ and the N\'eel temperature $T_{\text{N}}$. For
temperatures above $T_0$, the resistivity components show the negative
temperature coefficient characteristic of the single-ion Kondo effect,
whereas for $T$ less than $T_0$ the temperature coefficient changes to
positive, even though in the disordered samples (for $0<x<1$) the
resistivity has a finite zero-temperature limit. Superposed on this
background is an abruptly appearing change for $T<T_{\text{N}}$ caused by the
growing magnetic order as discussed above. Here one has to keep in
mind that the coherence temperature by definition can not be uniquely
defined, as it is meant to characterize a smooth crossover rather than
a phase transition.  The positions of the maxima of $\rho$, which are
significant for the onset of coherence, are seen to vary broadly for
the different resistivity components, reflecting the complex band
structure of the system.  Nonetheless, the main features of the
resistivity fit well into this picture.  The data seem to indicate
that for $x \ge 0.5$ the system is in the regime where $T_0<T_{\text{N}}$,
while for $x\le 0.3$ we observe $T_0\gtrsim T_{\text{N}}$. This may in part be
responsible (beside the much smaller magnetization for $x\le 0.3$
suppressing all quadratic effects) for the fact that the decrease of
resistivity perpendicular to $\Q$ is much larger for the higher Au
concentrations.

\section{Conclusions}
The neutron-scattering data of the heavy-fermion alloy
CeCu$_{6-x}$Au$_x$ reported here show that a complex magnetic
structure appears at low temperatures. As discussed in previous
publications, near the quantum-critical point at $x \approx 0.1$
two-dimensional fluctuations are found to dominate.  In the
magnetically ordered phase for $x>0.1$, considered in this paper,
incommensurate order emerges, characterized by an ordering wave vector
in the $a^*c^*$ plane for $x=0.2$ and $0.3$, changing to a different
wave vector for $x=0.5$ and $x=1$.  The appearance of magnetic order
has a profound effect on the electrical-resistivity components.
Broadly speaking, the resistivity tends to increase with staggered
magnetization for current directions along the wave vector $\Q$, while
it tends to decrease for all other directions.  between $x=0.3$ and
$x=0.5$. We have shown that a pronounced increase of the resistivity
proportional to the staggered moment is indeed expected for current
parallel to $\Q$. This effect is caused by the change in the band
structure induced by the scattering of the heavy quasiparticles off the
periodically varying magnetization. In addition we identified two
contributions in the resistivity proportional to the staggered moment
squared, the first one induced by the partial quenching of the Kondo
effect through the staggered magnetic field and therefore negative,
and the second one following from the change of the momentum
relaxation rate, with positive (negative) sign for $Q>2 \kf$ ($Q< 2
\kf$).  We found indications that $Q$ is close to $2 \kf$, where $\kf$
is a Fermi vector of the band structure in the direction of $\Q$.

In this scenario we obtain a satisfactory  picture
of the interplay of magnetic order and transport in CeCu$_{6-x}$Au$_x$
for $0.2 \le x \le 1$.
A more detailed understanding requires knowledge of the electronic
band structure of CeCu$_6$, as well as a microscopic theory of the
Kondo lattice and to the heavy fermion states, which is not yet
available.

\begin{acknowledgement}
We wish to thank N. Pyka and S. Pujol from the Institut Laue-Langevin
in Grenoble and E. Garcia-Matres, R. v. d. Kamp, S. Welzel and
H. Schneider at the Hahn-Meitner-Institut
 in Berlin for their help on neutron scattering and 
G. Portisch and H.~G.~Schlager for  preparing some of the crystals.
We acknowledge the assistance of R.~H\"au{\ss}ler
on some resistivity measurements.
This work was supported by the Deutsche Forschungsgemeinschaft.
\end{acknowledgement}

\end{document}